\documentclass[11pt]{elsarticle}
\journal{Nuclear Physics A}

\usepackage{graphicx,color,amsmath,amssymb}
\usepackage{bm}
\usepackage{hyperref}

\begin{document}

\begin{frontmatter}
\title{Fluctuation induced equality of multi-particle eccentricities for four or more particles}

\author[rbrc]{Adam Bzdak}
\ead{abzdak@bnl.gov}
\author[agh,inp]{Piotr Bozek}
\ead{piotr.bozek@ifj.edu.pl}
\author[bnl,rbrc,ccnu]{Larry McLerran}
\ead{mclerran@bnl.gov}

\address[rbrc]{RIKEN BNL Research Center, Brookhaven National Laboratory, Upton NY 11973, USA}

\address[agh]{AGH University of Science and Technology, Faculty of Physics and Applied Computer
Science, al. Mickiewicza 30, 30-059 Krakow, Poland}
\address[inp]{Institute of Nuclear Physics PAN, 31-342 Krak«ow, Poland}

\address[bnl]{Physics Dept, Bdg. 510A, Brookhaven National Laboratory, Upton, NY-11973, USA}
\address[ccnu]{Physics Dept, China Central Normal University, Wuhan, China}

\begin{abstract}
We discuss eccentricities (ellipticity and triangularity) generated in 
nucleus-nucleus and proton-nucleus collisions.  We define multi-particle eccentricities
$\epsilon_n\{m\}$ which are associated with the $n'th$ angular multipole moment for $m$ particles.  
We show that in the limit of
fluctuation dominance all of the $\epsilon_n\{m\}$'s are approximately equal for $m \ge 4$.
For dynamics linearly responding to these eccentricities such as hydrodynamics or proposed in this paper 
weakly interacting field theory, these relations among eccentricities are translated into 
relations among flow moments $v_n\{m\}$. 
We explicitly demonstrate it with hydrodynamic calculations.
\end{abstract}
\end{frontmatter}


\section{Introduction}

In nucleus-nucleus (A+A) and proton-nucleus (p+A) collisions, there is an approximately boost invariant structure 
associated with an angular asymmetry of the two and many particle
correlation functions. In heavy ion collisions, this asymmetry is conventionally associated with hydrodynamic flow driven by angular asymmetries of the underlying
matter distribution \cite{Ollitrault:1992bk}. This angular asymmetry 
contains a component that is due to fluctuations in the transverse positions of particle interactions 
\cite{Alver:2006wh,Alver:2010gr}, and a component associated with source asymmetry 
at finite impact parameter of the collision. In p+A collisions, a variety of mechanisms leading to angular correlations 
have been proposed, 
some involving hydrodynamic like
scenarios as in A+A interactions \cite{Bozek:2011if} and some involving non-trivial angular correlations associated with the emission process \cite{Dusling:2013oia}. 

To quantify the momentum space distribution of particles, one may identify an angular harmonic of the momentum space distribution as
\begin{equation}
    v_n = {1 \over N} \int d^2p_T~ e^{in\phi} {{dN} \over {dyd^2p_T}}
\label{eq:vn}
\end{equation}
where $N = \int d^2p_T~{{dN} \over {dyd^2p_T}}$.
This is a complex quantity and for each event is of the form
\begin{equation}
  v_n = \eta_n e^{i\gamma_n}
\end{equation}
The angle $\gamma_n$ describes the orientation of the flow vector relative to some chosen coordinate axis, and $\eta_n$ is its modulus. When averaging over events, it must be true that $\langle v_n  \rangle = 0$ by rotational invariance.
  
Borghini, Dinh and Ollitrault introduced multi-particle correlations that measure the rotationally invariant part of the flow \cite{Borghini:2000sa}. For example, the two particle correlation is
\begin{equation}
     v_n^2\{2\} = \langle e^{in(\phi_1-\phi_2)} \rangle = \langle |v_2|^2 \rangle
\end{equation}
In a collision at fixed impact parameter, it is conventionally believed that this expectation 
value contains a piece associated with the geometry of the collision,
which if the impact parameter is sufficiently well defined is non fluctuating, plus a fluctuating component.
Originally these authors introduced higher order components associated with 4, 6 and more particle correlations 
to reduce contributions from two (four or more) particle non-flow correlations.
For example
\begin{equation}
    v_n^4\{4\} = 2\langle |v_n|^2 \rangle^2 - \langle |v_n|^4 \rangle
\end{equation}    
and
\begin{equation}
    v_n^6\{6\} = {1 \over 4 }\left[ \langle | v_n |^6 \rangle - 9 \langle | v_n |^2 \rangle %
    \langle| v_n |^4 \rangle + 12 \langle | v_n |^2 \rangle^3
    \right]
\end{equation}
and higher order generalizations to larger number of particles. Interesting for the following discussion is the fact, that  the cumulant expressions for the harmonic flow coefficients allow
one to isolate the fluctuating component from that of the geometry of the collision.
If there is no mean field contribution to the distribution functions then, it is easy to see that for purely Gaussian fluctuations,
$v_n\{4\} = v_n\{6\} = 0$, that is these terms are sensitive only to correlations of fourth order or higher from the average flow \cite{Voloshin:2007pc}. For a large number of independent sources, these correlations are expected to be small. So it is believed that these higher order correlations capture the intrinsic flow contributions and reduce the effects of fluctuations.

In heavy ion collisions it is in fact found that with very good precision $v_2\{4\} = v_2\{6\} = v_2\{8\}$ , with $v_2\{2\}$ different due 
to its intrinsic fluctuations \cite{Bilandzic:2012an,Chatrchyan:2013kba}. The pattern seen so far in p+A collisions is remarkably similar to that found in heavy ion collisions \cite{Aad:2013fja,Chatrchyan:2013nka}.
Why is this result surprising? It is because if one measures the magnitude of flow fluctuation by
$\sqrt{{{v_2\{2\}^2 - v_2\{4\}^2} \over {v_2\{2\}^2 + v_2\{4\}^2}}}$,
one finds that this is comparable in p+A and A+A collisions independent of impact parameter.
This would suggest that a mean field value is not 
so well defined  \cite{Aad:2013xma}. A related observation is that the measured value of $v_3\{4\}$ in A+A collisions deviates from zero
\cite{Adamczyk:2013waa,ALICE:2011ab,Chatrchyan:2013kba}. The triangularly in A+A interaction region comes only from 
fluctuations and one expects $v_3\{4\}\simeq 0$ in the leading order. However, for fluctuation dominated 
eccentricities, subleading effects give $\epsilon_3\{4\}\neq 0$ \cite{Bhalerao:2011yg}, which can 
explain the experimental data.

In this paper, we will consider in detail the generalization of $v_n\{m\}$ by introducing coefficients 
$\epsilon_n\{m\}$ ($m \ge 2$ is even). The advantage of this is that we can easily
compute the distribution of eccentricities, and understand the various contributions arising from a mean field or from fluctuations. Of course these quantities are the input into a computation
of the $v_n\{m\}$, but for small $\epsilon_n$ we expect linearity in the response to the system for the elliptic 
and triangular flow \cite{Gardim:2011xv,Niemi:2012aj}
\begin{equation}
   v_n\{m\}  = c_n \epsilon_n\{m\}
\label{eq:c_n}
\end{equation}
Here $c_n$ is independent of $m$. This result is well known for hydrodynamics but we also demonstrate it in the case where the source of eccentricity is radiation of a weakly coupled scalar field. We expect that this linear response is quite general.
     
We can define the eccentricities as
\begin{equation}
   \epsilon_n = {1 \over {\langle r_T^{n} \rangle}} \int d^2r_T e^{in\phi} r_T^{n} {{dN} \over {dyd^2r_T}}
\end{equation}      
The quantities $\epsilon_n\{m\}$ are defined in precise analogy to the $v_n\{m\}$.
 
Now we can state the main result of this paper. 
Performing numerical calculations we observe that 
$\epsilon _{n}\{2\}>\epsilon _{n}\{4\}\simeq \epsilon _{n}\{6\}\simeq \epsilon
_{n}\{8\}$, $n=2,3$, for both A+A and p+A collisions. Experimental observation of analogous relation 
for the Fourier coefficients of the azimuthal correlation function in p+A, $v_n\{m\}$, would indicate 
the importance of the initial geometry in such collisions as present, e.g., in hydrodynamics or 
weakly interacting field theory (see Section 3). 

It is worth mentioning that equality of multi-particle eccentricities may be obtained by a two-parameter
shifted Gaussian weight function
\begin{equation}
    \langle | \epsilon_n | ^{m} \rangle ={{ \int dz d \overline z e^{-(z-z_n)(\overline z - \overline z_n)/\sigma_n^2 } (z\overline z)^{m/2} } \over {\int dz d \overline z e^{-(z-z_n)(\overline z - \overline z_n)/\sigma_n^2}}}
\end{equation}
We use complex notation, $z = x+iy$. The result we find for this distribution is that
\begin{equation}
  \epsilon_n\{2\} = \sqrt{| z_n |^2 + \sigma_n^2 }
\end{equation}
\begin{equation}
  \epsilon_n\{m\} = | z_n |,
\end{equation}
where $m \ge 4$. Certainly there are different distributions leading to an approximate equality of
$\epsilon_n\{m\}$ for $m \ge 4$. We will come back to this point at the end of this paper.

Some of the discussion we present in this paper is not new, and has been derived in some form in the literature 
referred to in this paper. Our goal here is to tie together various common features of p+A and A+A collisions, 
and in particular the appearance of a mean
ellipticity and triangularity even in the fluctuation dominated region of such collisions.  
We hope that the observations we present will help to focus the understanding of contrasting explanations 
of p+A and A+A collisions.

\section{Multi-particle eccentricities}

\begin{figure}[h!]
\begin{center}
\includegraphics[scale=0.35]{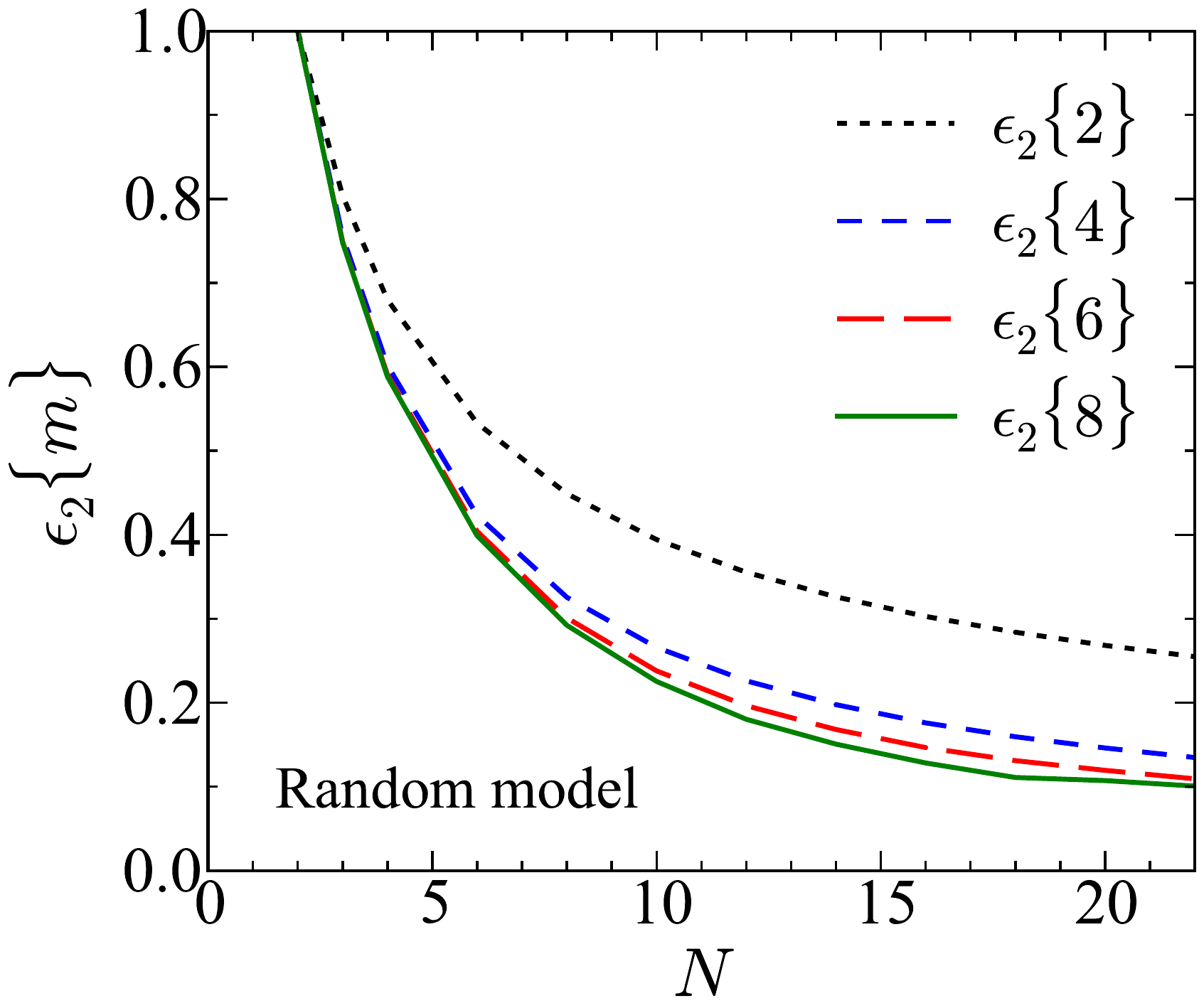} %
\includegraphics[scale=0.35]{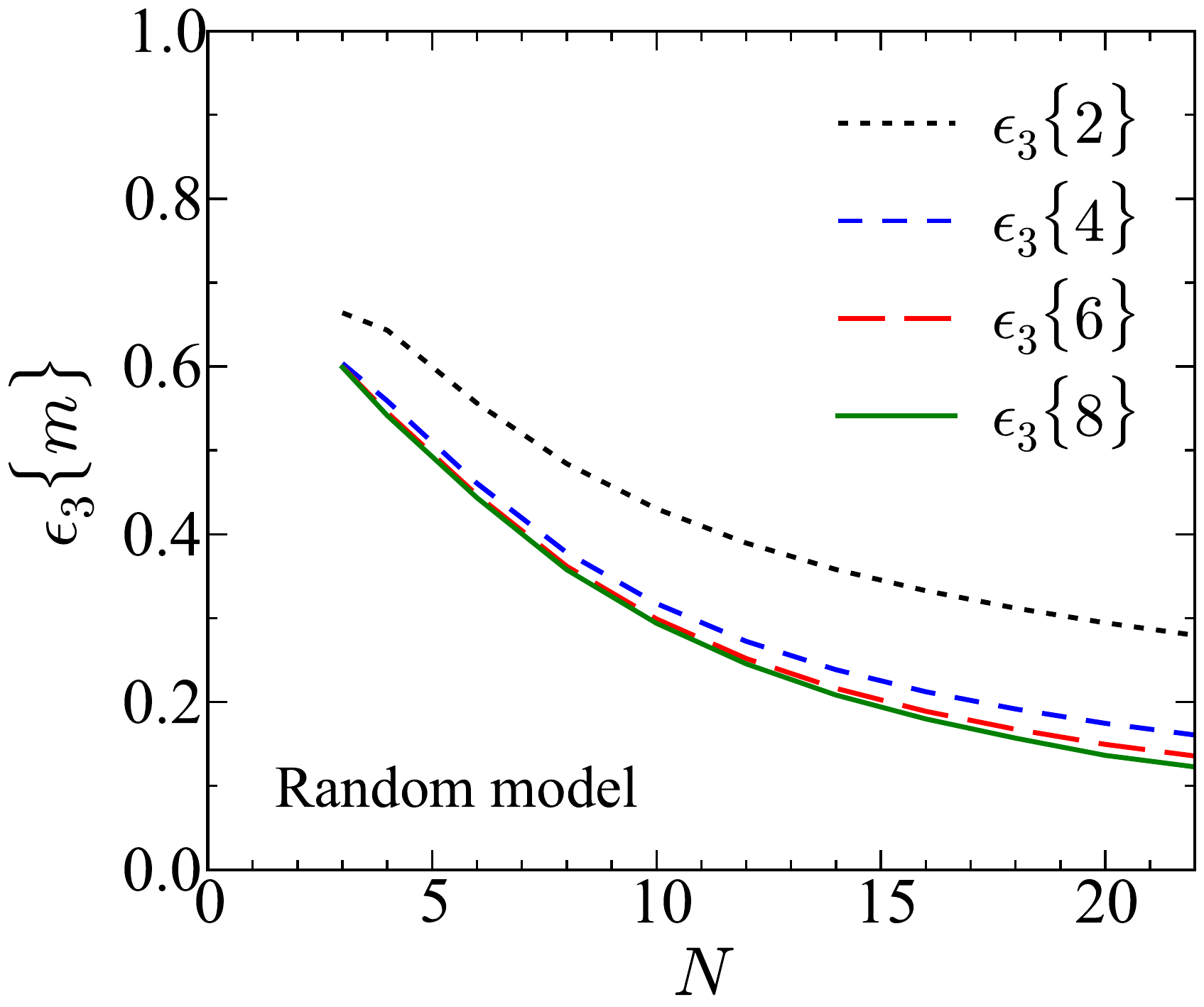}
\end{center}
\caption{The $m$-particle eccentricities $\epsilon _{n}\{m\}$, $%
m=2,4,6,8$ for $n=2$ (left) and $n=3$ (right) calculated in the random model
for a various number of points, $N$, sampled randomly on the two dimensional
plane.}
\label{fig:rand:1}
\end{figure}

We start our discussion with a simple model with a given number of points
sampled randomly on the two dimensional plane. In this model we take $N$
points with the $x$ and $y$ coordinates sampled from a uniform distribution
in the interval $[-1,1]$ and calculate $\epsilon_{n}\{m\}$ for all points
satisfying $x^{2}+y^{2}<1$. In each event we obtain 
(the coordinate system is shifted to the center of mass)
\begin{equation}
\epsilon _{n}^{2}=\frac{\left[ \sum_{i=1}^{N}r_{i}^{n}\cos (n\phi _{i})%
\right] ^{2}+\left[ \sum_{i=1}^{N}r_{i}^{n}\sin (n\phi _{i})\right] ^{2}}{%
\left[ \sum_{i=1}^{N}r_{i}^{n}\right] ^{2}},
\label{eq:epsdef}
\end{equation}%
where $r_{i}^{2}=x_{i}^{2}+y_{i}^{2}$, and finally we average over the
sufficient number of events. In Fig. \ref{fig:rand:1} we present our results
for $\epsilon _{n}\{m\}$, $m=2,4,6,8$ and $n=2,3$.\footnote{%
For completeness we have (for $\epsilon _{n}\{2,4,6\}$ see Section 1):%
\begin{eqnarray*}
33\epsilon _{n}\{8\}^{8}&=&-\langle \epsilon _{n}^{8}\rangle
+16\langle \epsilon _{n}^{6}\rangle \langle \epsilon
_{n}^{2}\rangle +18\langle \epsilon _{n}^{4}\rangle
^{2}-144\langle \epsilon _{n}^{4}\rangle \langle \epsilon
_{n}^{2}\rangle ^{2}+144\langle \epsilon _{n}^{2}\rangle
^{4}
\end{eqnarray*}
}
As discussed in the introduction it is interesting to notice that%
\begin{equation}
\epsilon _{n}\{2\}>\epsilon _{n}\{4\}\simeq \epsilon _{n}\{6\}\simeq \epsilon
_{n}\{8\},\text{\quad }n=2,3  \label{eps-relations}
\end{equation}
Expanding the expectations in Eq. (\ref{eq:epsdef}) to a leading order one finds a Bessel-Gaussian distribution  
with zero mean-field
for the event-by-event distribution of $\epsilon_{n}$ \cite{Broniowski:2007ft,Voloshin:2007pc}, for 
which $\epsilon _{n}\{m\}=0$
for $m\ge 4$. 
Our results show that $\epsilon_n\{m\}$ for $m\ge 4$ are non zero, and for the values of $N$ that we consider
are roughly of the same order of magnitude as the width of the splitting between the second moment and all of the others (Fig. \ref{fig:rand:1}).  The moments for $m \ge 4$ are to a good
approximation equal. This means that the deviations from the Bessel-Gaussian limit are roughly
of the same order as the fluctuations giving a nonzero value of 
$\langle |\epsilon_n|^2 \rangle$
 \cite{Bhalerao:2006tp,Alver:2008zz,Bhalerao:2011yg}.
\begin{figure}[h!]
\begin{center}
\includegraphics[scale=0.35]{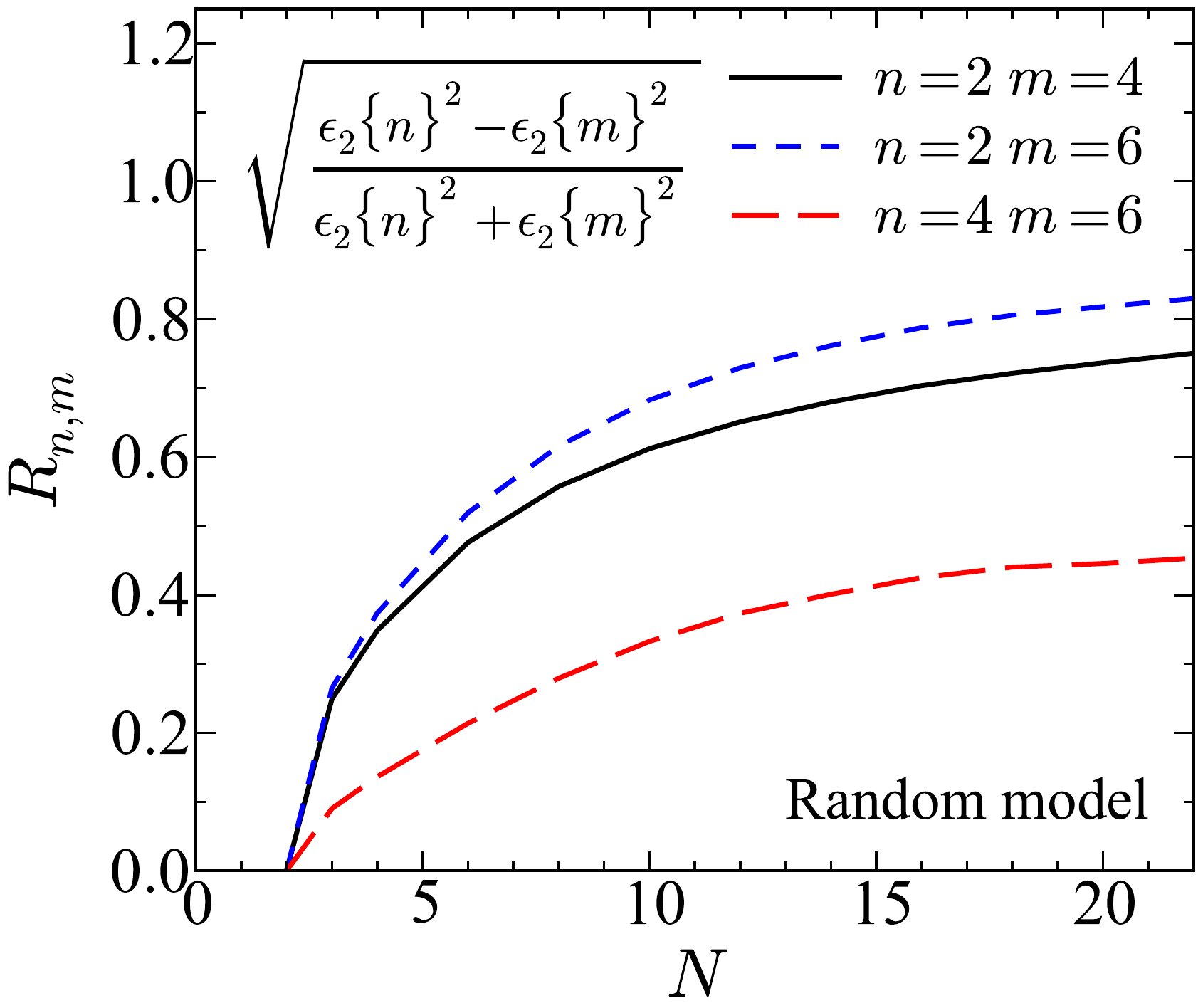}
\includegraphics[scale=0.35]{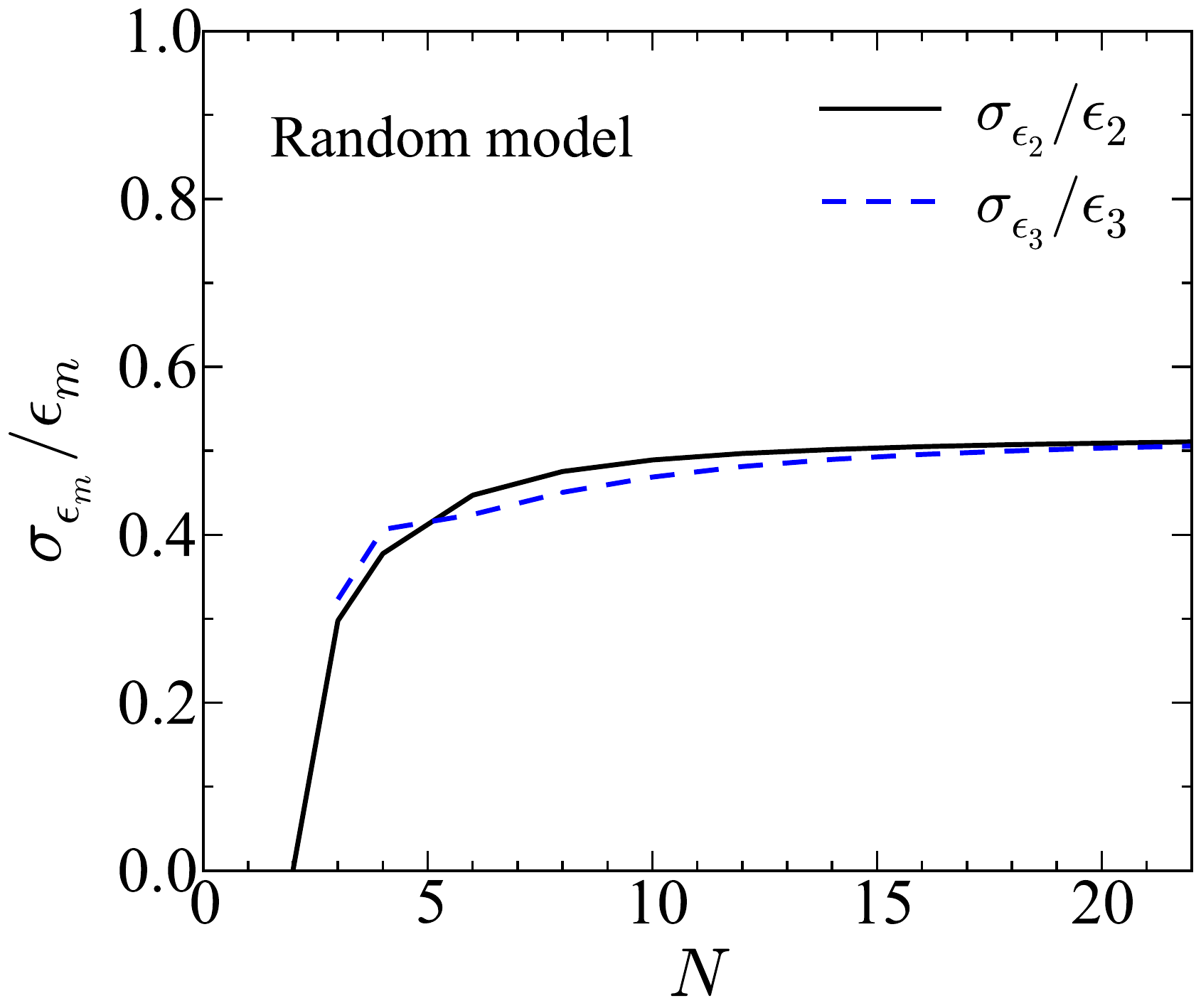}
\end{center}
\caption{The ratio $R_{n,m}$ (left), defined in the plot, and the width of $\epsilon _{n}$ distribution over the average
value (right), $[ \langle \epsilon _{n}^{2}\rangle
-\langle \epsilon _{n} \rangle ^{2}]
^{1/2}/\langle \epsilon _{n} \rangle $, calculated in the
random model for various number of points, $N$, sampled randomly on the two
dimensional plane.}
\label{fig:rand:2}
\end{figure}

In the left panel of Fig. \ref{fig:rand:2} we evaluate 
\begin{equation}
    R_{n,m} = \sqrt{{{\epsilon_2\{n\}^2 - \epsilon_2\{m\}^2} \over {\epsilon_2\{n\}^2 + \epsilon_2\{m\}^2}}}
\end{equation}
for $n=2,4$ and $m=4,6$. This quantity allows for a detail comparison of the mean field and fluctuating components. 
In the right panel of Fig. \ref{fig:rand:2} we present
\begin{equation}
 \frac{\sigma_{\epsilon _{n}}}{\left\langle \epsilon _{n}\right\rangle} = 
 \frac{\sqrt{ \left\langle \epsilon_{n}^{2}\right\rangle -\left\langle \epsilon _{n}\right\rangle ^{2}}}
      {\left\langle \epsilon _{n}\right\rangle}
\end{equation}
which measures the
strength of the fluctuations with respect to the average value of $%
\left\langle \epsilon _{n}\right\rangle $. Already for $N>10$ it approaches
the limit $\sqrt{-1+4/\pi }$, derived in Ref. \cite{Broniowski:2007ft}. 

\begin{figure}[h!]
\begin{center}
\includegraphics[scale=0.35]{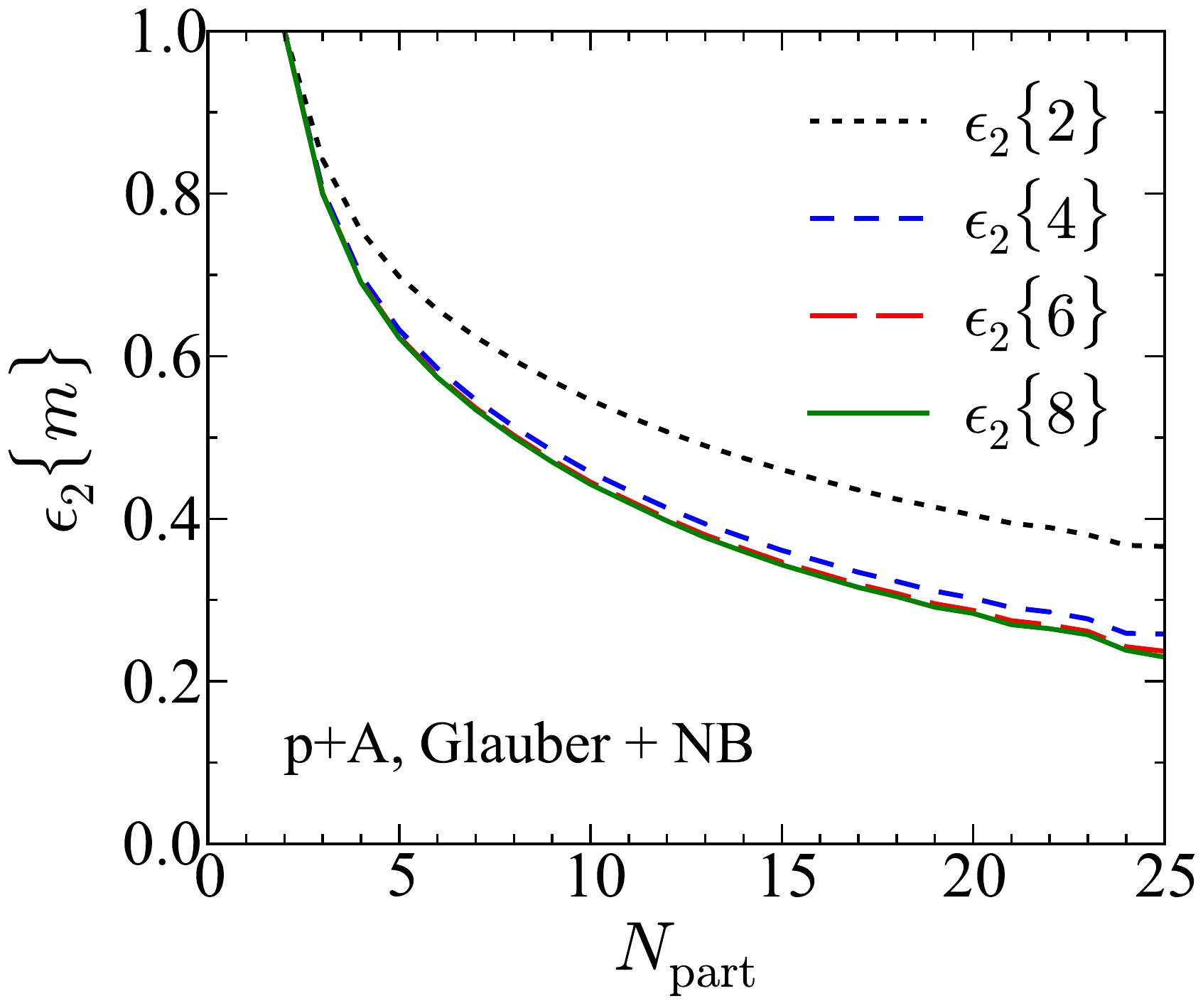} %
\includegraphics[scale=0.35]{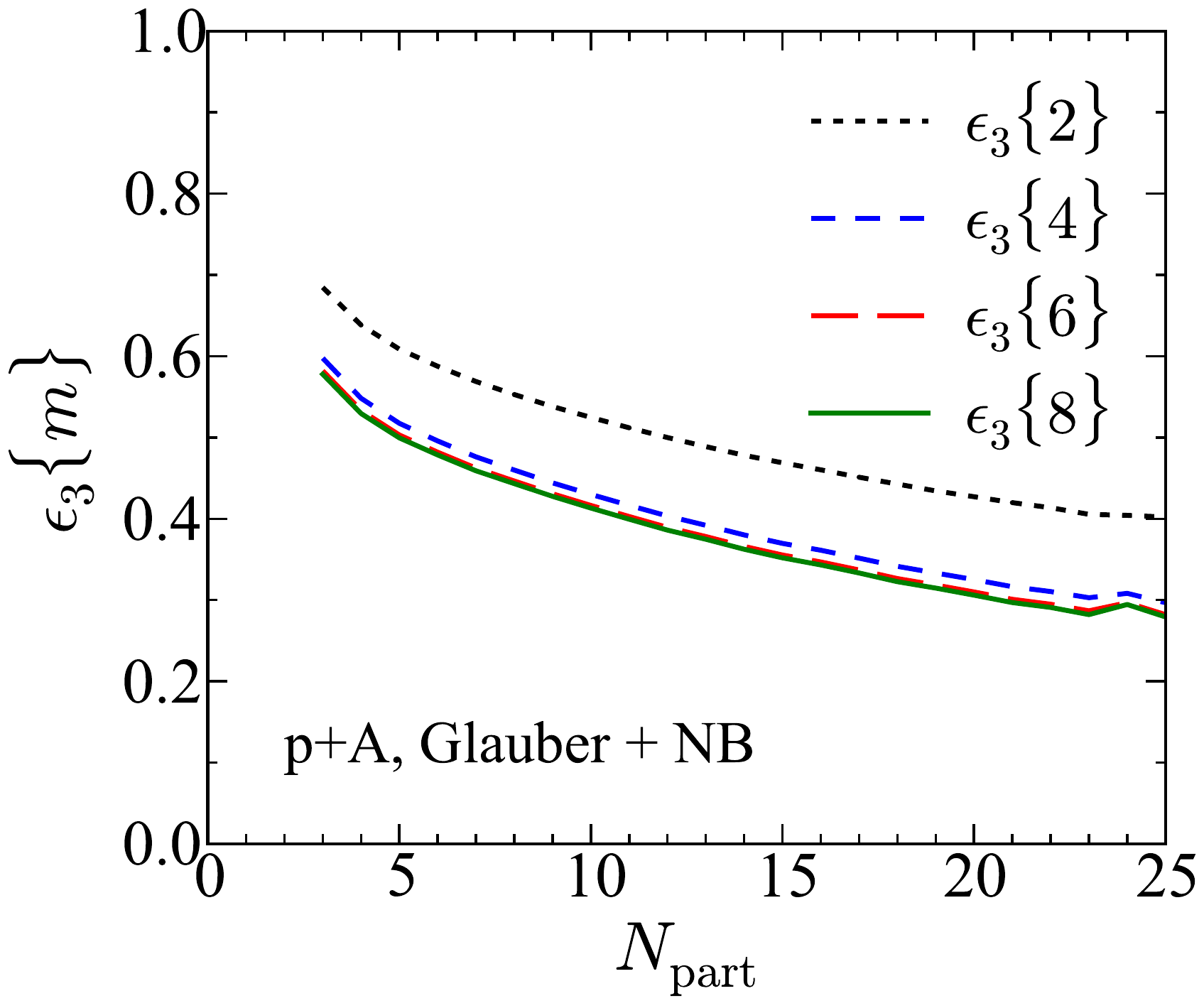}
\end{center}
\caption{The $m$-particle eccentricities $\epsilon _{n}\{m\}$, 
$m=2,4,6,8$ for $n=2$ (left) and $n=3$ (right) versus the number of wounded
nucleons, $N_{\rm part}$, calculated in the Glauber model for p+Pb with an additional fluctuations, in
the centers of participants, given by the Gamma distribution.}
\label{fig:pA:1}
\end{figure}

It is obvious that the random model discussed above is driven only by
fluctuations. It suggests that identical properties for $\epsilon _{n}\{m\}$,
in particular Eq. (\ref{eps-relations}), should be also present in
p+A collisions. Indeed, performing suitable calculations in the
standard Glauber model, see e.g., \cite{Rybczynski:2013yba} we obtain very similar results, both
qualitatively and quantitatively, to those presented in Fig. \ref{fig:rand:1}. 
We checked that this conclusion is independent on the
specific realization of the Glauber model. For example, in Fig. \ref{fig:pA:1} 
we show the results based on the Glauber model in p+Pb collisions with additional
fluctuations, given by the Gamma distribution, deposited at the positions of the wounded
nucleons \cite{Bozek:2013uha}. The additional fluctuations of the 
source lead to increase
of eccentricities for higher cumulants, e.g., $\epsilon_2\{4\}/\epsilon_2\{2\}\simeq 0.7$.
For very large values of $N$ we would expect deviations from
this expectation \cite{Alver:2008zz,Bhalerao:2011yg}, with $\epsilon_2\{4\}$ decreasing as $1/N^{3/4}$, while 
$\epsilon_2\{2\}$ behaves as $1/N^{1/2}$.
\begin{figure}[h!]
\begin{center}
\includegraphics[scale=0.35]{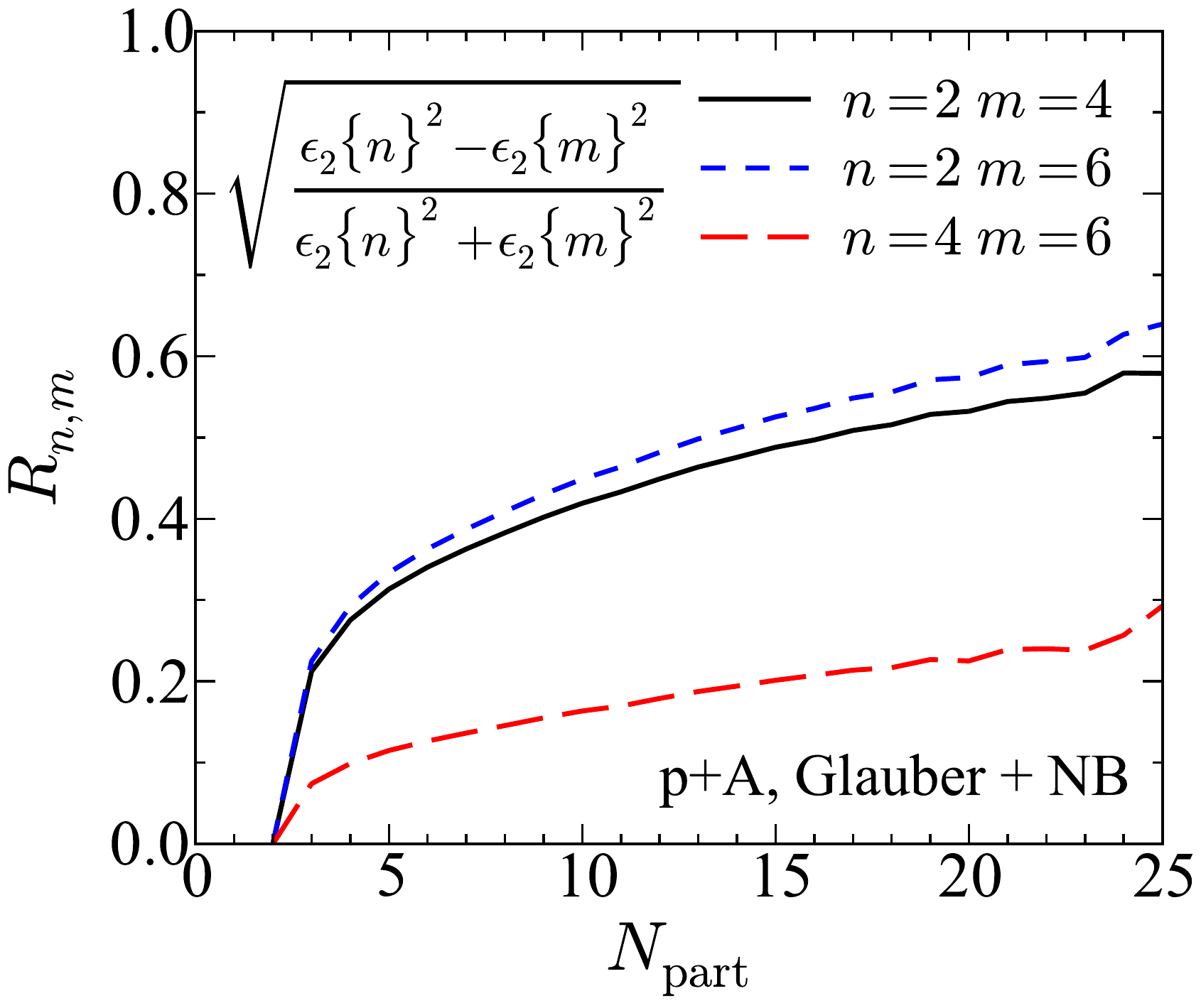}
\includegraphics[scale=0.35]{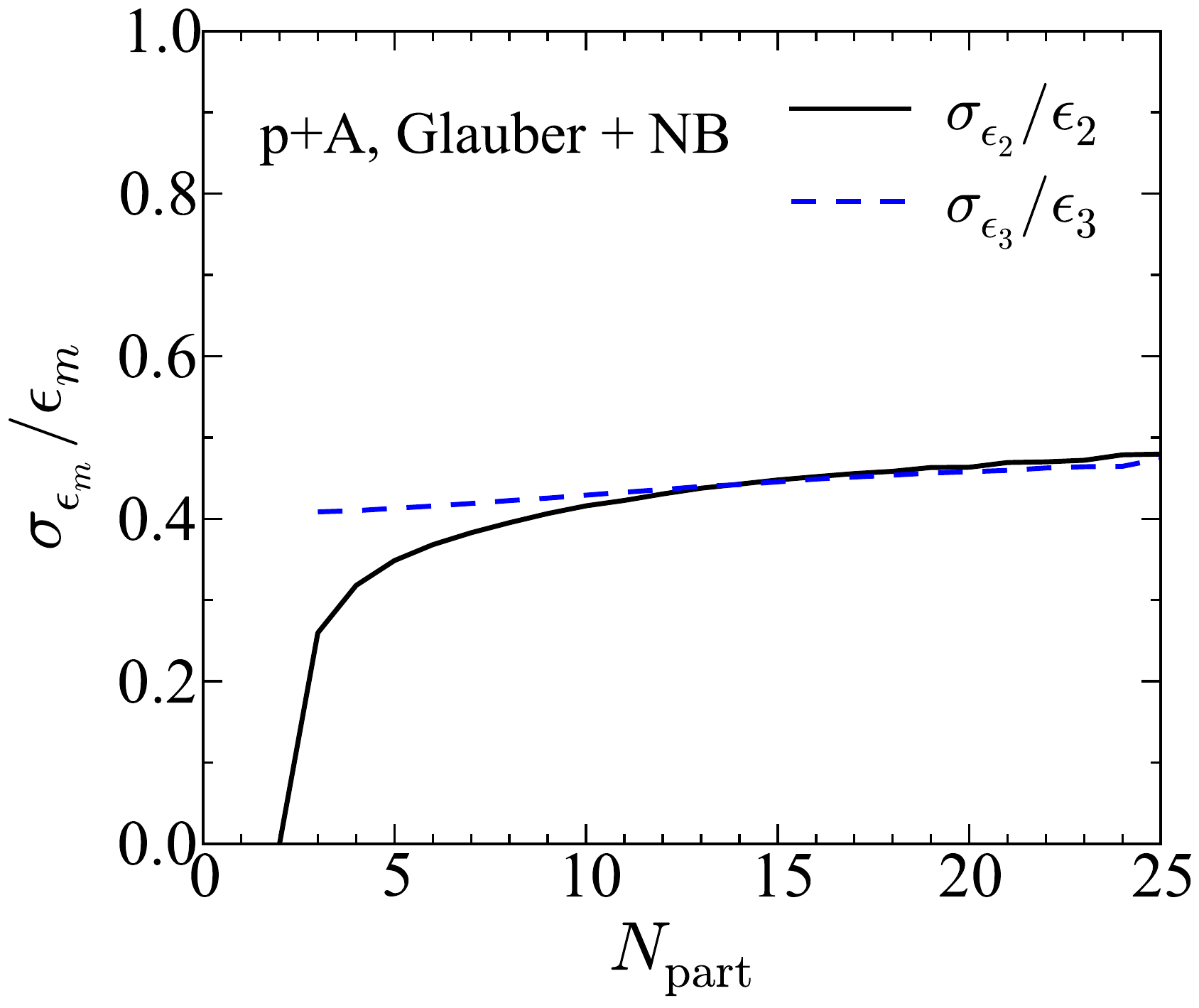}
\end{center}
\caption{The ratio $R_{n,m}$ (left), defined in the plot, and the width of $\epsilon _{n}$ 
distribution over the average
value (right), $[ \langle \epsilon _{n}^{2}\rangle
-\langle \epsilon _{n} \rangle ^{2}]
^{1/2}/\langle \epsilon _{n} \rangle $ versus the number
of wounded nucleons, $N_{\rm part}$, calculated in the Glauber model for p+Pb with an additional
fluctuations in the center of participants given by the Gamma distribution.
}
\label{fig:pA:2}
\end{figure}

The obtained signal satisfies the same relation as in Eq. (\ref%
{eps-relations}). In Fig. \ref{fig:pA:2} we present results for $R_{n,m}$ (left panel) and 
$\sigma_{\epsilon _{n}} / \langle \epsilon _{n} \rangle$ (right panel). In both figures we plot as a 
function of the number of wounded nucleons \cite{Bialas:1976ed} coming form the Glauber calculation.

\begin{figure}[h]
\begin{center}
\includegraphics[scale=0.35]{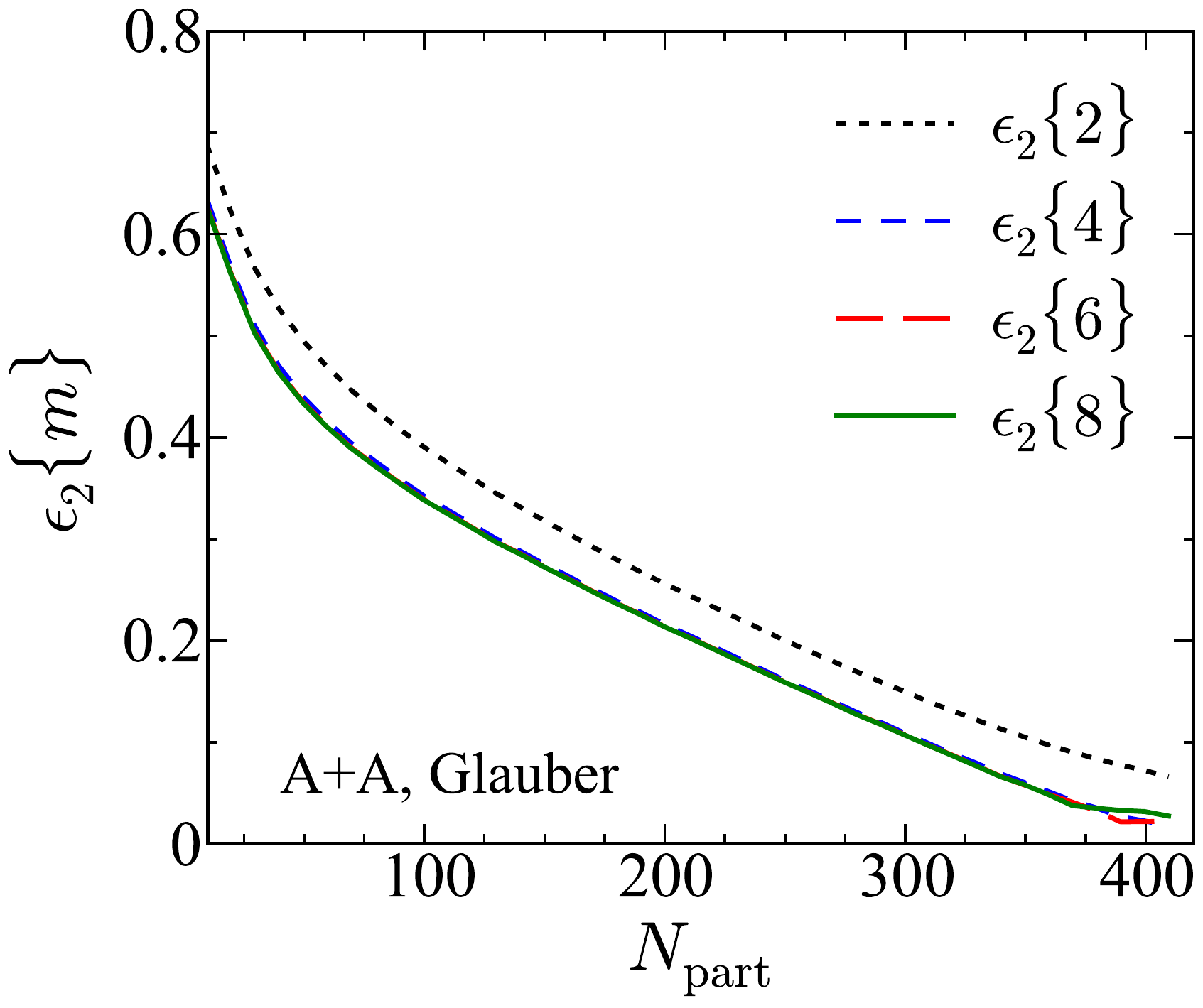} %
\includegraphics[scale=0.35]{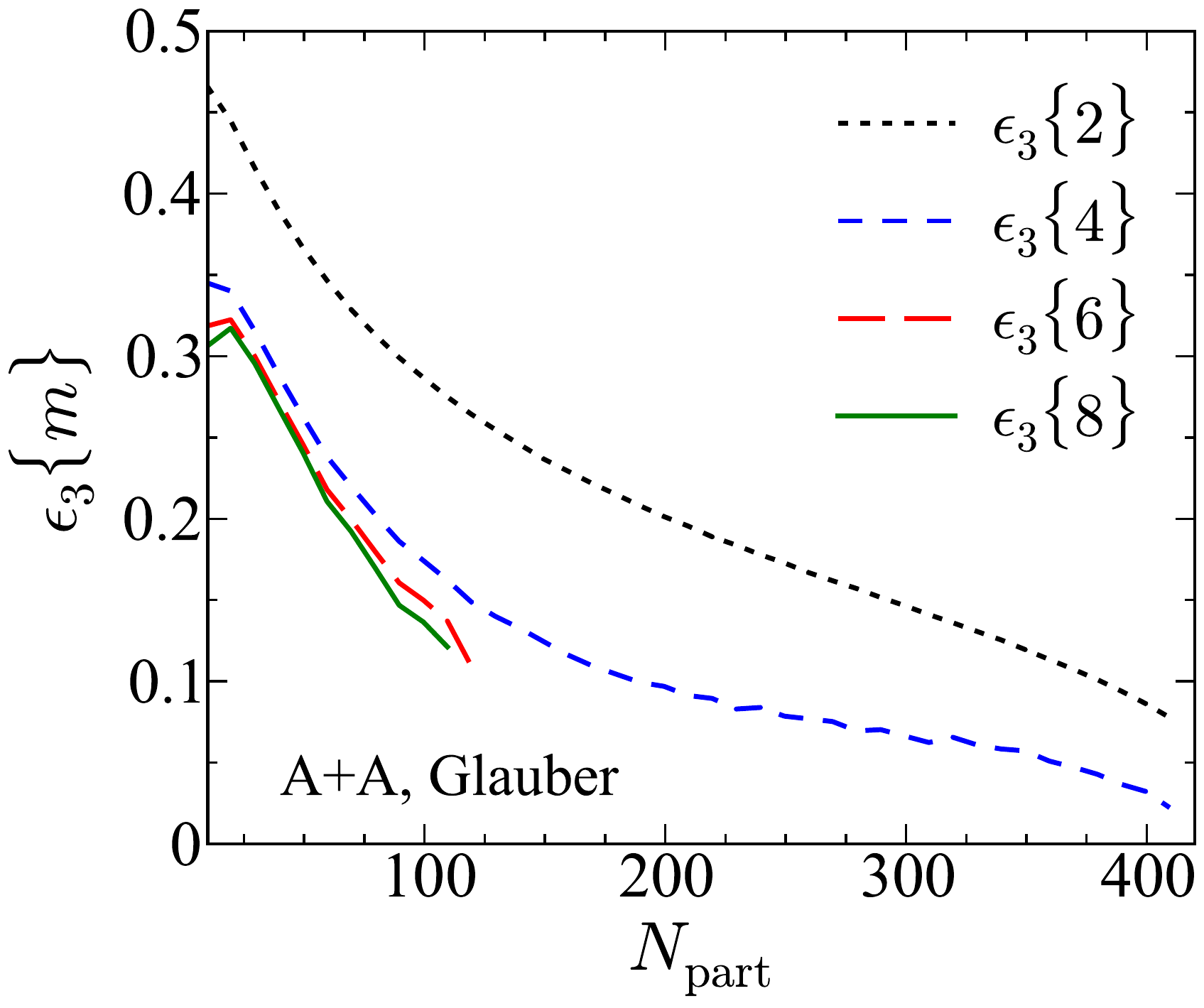}
\end{center}
\caption{The $m$-particle eccentricities $\epsilon _{n}\{m\}$, 
$m=2,4,6,8$ for $n=2$ (left) and $n=3$ (right) versus the number of wounded
nucleons, $N_{\rm part}$, calculated in the standard Glauber model for Pb+Pb collisions. 
We cut $\epsilon _{3}\{6,8\}$ for $N_{\rm part}>110$ owing to insufficient statistics. }
\label{fig:AA:1}
\end{figure}

\begin{figure}[h]
\begin{center}
\includegraphics[scale=0.35]{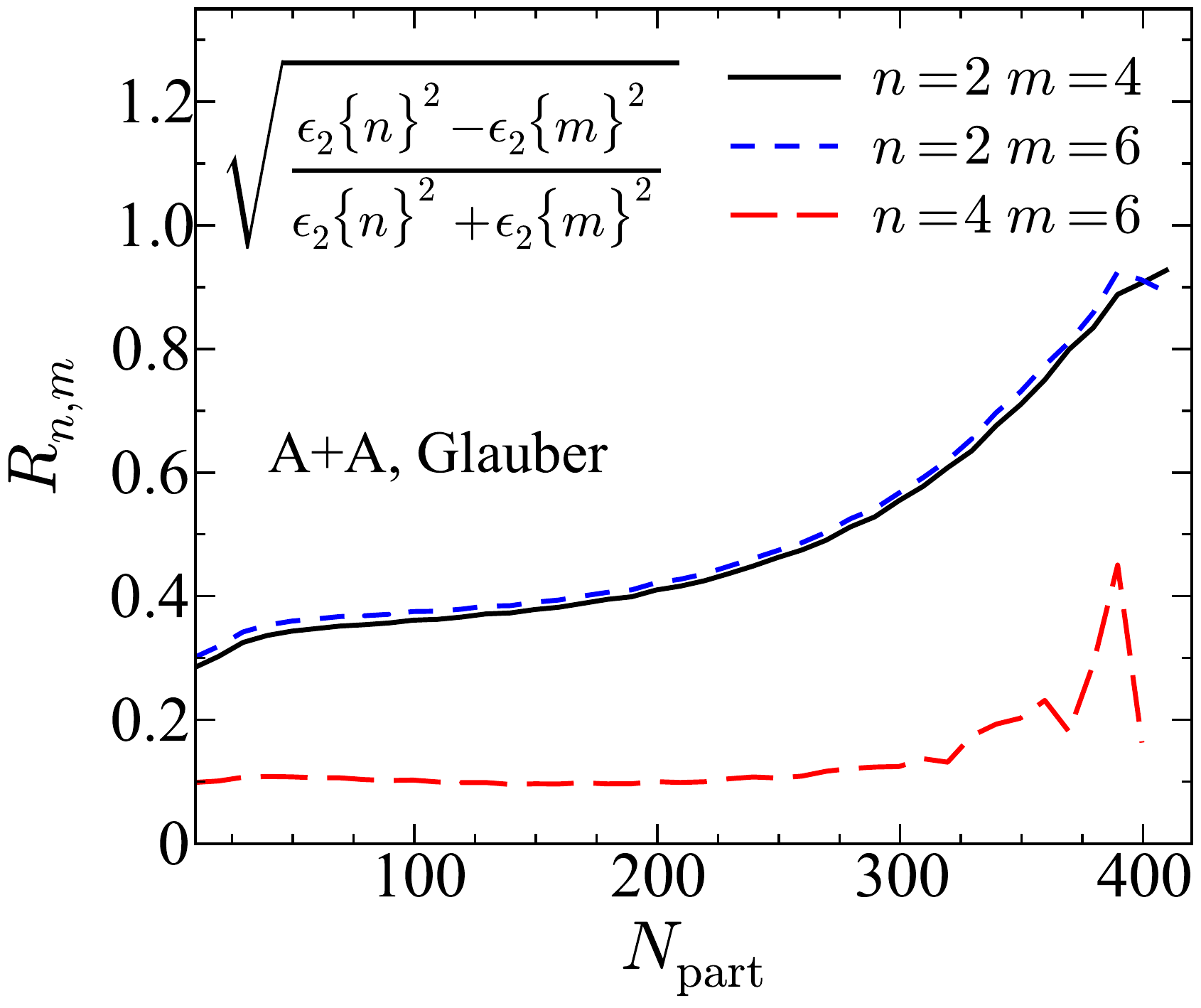}
\includegraphics[scale=0.35]{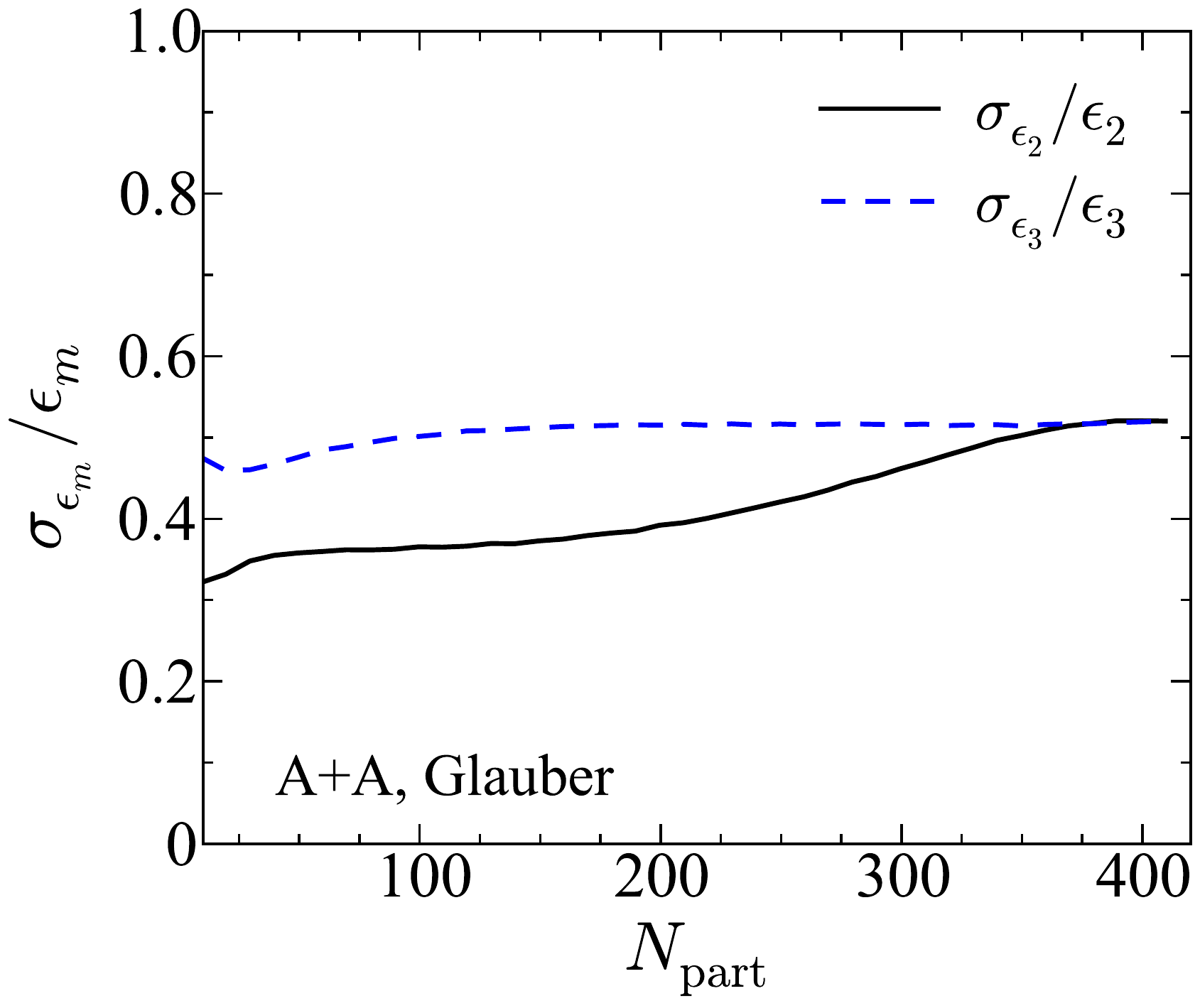}
\end{center}
\caption{The ratio $R_{n,m}$ (left), defined in the plot, and the width of $\epsilon _{n}$ 
distribution over the average
value (right), $[ \langle \epsilon _{n}^{2}\rangle
-\langle \epsilon _{n} \rangle ^{2}]
^{1/2}/\langle \epsilon _{n} \rangle $ versus the number
of wounded nucleons, $N_{\rm part}$, calculated in the standard Glauber model for Pb+Pb
collisions.}
\label{fig:AA:2}
\end{figure}

At the end of this Section we present in Fig. \ref{fig:AA:1} results for Pb+Pb collisions.
Ellipticity in A+A collisions is not only driven by fluctuations but also a
mean field component is present in an off central collisions. It is clear that if eccentricities are 
dominated by a mean field we expect 
$\epsilon _{n}\{2\}\simeq\epsilon _{n}\{4\}\simeq \epsilon _{n}\{6\}\simeq \epsilon_{n}\{8\}$ but, as we argue, it is never the case in A+A collisions. As shown in Fig. \ref{fig:AA:2}, the quantity $R_{2,4}$
and the scaled width of the fluctuations are actually comparable to p+A collisions,
which are solely driven by fluctuations. Taking into account our discussion of p+A collisions it is 
now obvious that in A+A collisions the relation 
$\epsilon _{n}\{2\}>\epsilon _{n}\{4\}\simeq \epsilon _{n}\{6\}\simeq \epsilon
_{n}\{8\}$ holds as well for all impact parameters.

\section{Discussion}

The elliptic and triangular asymmetry in the collective flow 
arises from the corresponding elliptic and triangular deformations of the initial source. 
Event-by-event simulations show that the eccentricity
distributions in A+A collisions are reproduced in the harmonic flow distribution for $v_2$ and $v_3$
\cite{Schenke:2012hg,Gardim:2011xv,Niemi:2012aj}. The same correspondence has been noticed 
for p+Pb collisions \cite{Bozek:2013uha}, see also \cite{Bzdak:2013zma}. Advanced hydrodynamic calculations 
using IP-Glasma initial conditions reproduce quantitatively the event-by-event $v_n$ 
distributions \cite{Schenke:2012hg}.\footnote{Although the results were shown only for $0-5\%$ and $20-25\%$ 
centrality classes.} 
Calculations using Glauber or MC-KLN models cannot reproduce the experimental distributions 
of $v_2$ for all the centralities.
This is due mainly to differences of the predicted mean-field eccentricities which lead to 
$\epsilon_n$ distribution shapes that cannot be rescaled to fit the experimental distributions 
of $v_2$  \cite{Aad:2013xma}.
The situation is different for $v_2$ in ultracentral A+A and for $v_3$, the measured distributions are close to the
Bessel-Gaussian, and are compatible with rescaled $\epsilon_n$ distributions obtained in 
the Glauber and MC-KLN models \cite{Aad:2013xma}. Moreover, the numerical examples presented above, show that the approximate 
equality of higher cumulant eccentricities is independent of details of how the fluctuations are 
generated in the initial density, whenever the fluctuations dominate.

In p+A collisions the eccentricities are fluctuation dominated and the Glauber model is expected to give 
a qualitatively correct picture of eccentricity distributions. Our results show that the elliptic 
and triangular eccentricities flow coefficients for higher order cumulants are all of the same order 
for p+A collisions.
The eccentricities $\epsilon_n\{4\}\simeq\epsilon_n\{6\}\simeq\epsilon_n\{8\}$ are $60-70\%$ of $\epsilon_n\{2\}$, depending on the details of the Glauber model. From the approximately linear response of hydrodynamic flow to the  initial asymmetries, we expect 
a similar relation for $v_2\{m\}$ and $v_3\{m\}$. With an additional increase of $v_n\{2\}$ from
non-flow two-particle correlations.

From  event-by-event hydrodynamic 
simulations for centrality $0$-$10$\% using Glauber+NB (NB stands for negative binomial) initial 
conditions \cite{Bozek:2013uha} one obtains 
the  distribution of the elliptic flow coefficients $v_2$, without non-flow or
finite multiplicity fluctuations. The cumulants of the
elliptic flow coefficient
distribution  give 
\begin{eqnarray}
	v_{2}\{2\} &=& 0.082\pm 0.002,  \notag \\
	v_{2}\{4\} &=& 0.055\pm 0.004,  \notag \\
	v_{2}\{6\} &=& 0.052\pm 0.005.
\end{eqnarray}
The relation between different $v_2\{m\}$ is in fair agreement with the values of the initial $\epsilon_2\{m\}$.

One can also find a relationship of the type of Eq. (\ref{eq:c_n}), that expresses the independence of the flow 
induced by eccentricity to the
number of particles involved  making the correlation. This follows from free field theory and arises 
from the interference pattern
of waves emitted by sources corresponding to the eccentricity.  For example, suppose we have a source 
with a dipole moment that emits at the time of the collision. Then
\begin{equation}
    (k^2 +M^2) \phi(k) = \rho(k)
\end{equation}
If the source has a dipole moment in space it is transformed into a dipole moment in momentum space 
after Fourier transformation.
If one computes the distribution of the source squared, $|\rho(k)|^2$ one therefore has a quadruple 
moment corresponding to $\epsilon_2$.
That is the quadruple asymmetry of the source distribution $|\rho(r)|^2$ is transformed into 
a quadruple asymmetry in momentum space by 
wave interference. Since $v_2\{m\} \sim \langle | \phi |^{2} \rangle$, we gain the equality of 
the flow moments. Of course, we need to introduce
some fluctuation into the correlation with $v_2\{2\}$ in order to reproduce the pattern we see, which 
presumably arise from quantum fluctuations around the classical wave solution.  
We will critically discuss this possibility in later work.   

\section{Conclusions}

The  elliptic and triangular eccentricity distributions from fluctuations of finite number, $N$, of sources 
show deviations form the Bessel-Gaussian distribution. The eccentricities calculated from the 
2, 4, 6, or 8- particle cumulants
are nonzero. The second order eccentricity $\epsilon_n\{2\}$ is always 
the largest. Expressions with higher order
cumulants give similar values $\epsilon_n\{4\}\simeq\epsilon_n\{6\}\simeq\epsilon_n\{8\}$, both in p+A and A+A
collisions. Typically for the p+A system
$\epsilon_n\{m\}\simeq (0.6-0.7) \epsilon_n\{2\}$ for $m\ge 4$ .
The harmonic flow coefficients $v_n\{m\}$ undergo  similar relations, when hydrodynamic 
expansion translates the initial shape 
asymmetry into the azimuthal asymmetry of emitted particles. 
For collisions where fluctuations dominate the higher order cumulant, $v_n\{m\}$ are not zero for Glauber 
model initial conditions.
The nonzero value of $v_n\{6\}$ or $v_n\{8\}$ can arise solely from fluctuations due to a finite number 
of sources, and with explicit symmetry breaking in A+A collisions at finite impact parameter.
We also proposed a novel mechanism which linearly translates $\epsilon_n\{m\}$ into $v_n\{m\}$ without 
dynamical interaction between produced particles.   

In view of the above discussion, it would be interesting to study and compare the higher order cumulant 
correlations in p+A collisions and also to compare higher order cumulant results for $v_2$ and $v_3$ in 
ultracentral A+A collisions. 

\bigskip

After this paper was submitted to arXiv two papers appeared discussing similar problem.
In Ref. \cite{Yan:2013laa} it was noticed that the distribution of $\epsilon_{2}$ in p+A is well described by a power 
law function, which naturally leads to an approximate equality of multi-particle eccentricities 
$\epsilon_{2}\{m\}$ for $m \ge 4$. In Ref. \cite{Bzdak:2013raa} the relation between 
various cumulants $\epsilon_{2}\{m\}$'s, in particular an approximate equality of higher multi-particle 
eccentricities is calculated analytically.

\section*{Acknowledgments}

A.B. thanks Gabriel Denicol for interesting conversations. 
L.M. gratefully acknowledge very useful discussions with Sergei Voloshin and Arthur
Poskanzer concerning the seminal work that they pioneered in the study of flow and eccentricity cumulants.
L.M. also acknowledges an important discussion with Jean-Yves Ollitrault concerning the dependence of various cumulants
on the number of sources at large number of sources.   We also gratefully acknowledge a comment sent to us by Jurgen Schukraft when this
paper was being prepared for publication, showing numerical computations based on a Glauber model in substantial agreement
with our numerical results.
A.B. is supported through the RIKEN-BNL Research Center. 
P.B. is partly supported by the National Science Centre, Poland, grant DEC-2012/05/B/ST2/02528, and PL-Grid infrastructure.
The research of L.M. is supported under DOE Contract No. DE-AC02-98CH10886.


\end{document}